\documentclass[fleqn,usenatbib]{mnras}

\usepackage[T1]{fontenc}

\DeclareRobustCommand{\VAN}[3]{#2}
\let\VANthebibliography\thebibliography
\def\thebibliography{\DeclareRobustCommand{\VAN}[3]{##3}\VANthebibliography}

\usepackage{graphicx}	
\usepackage{amsmath}	
\usepackage{amssymb}	
\usepackage{scalefnt}

\usepackage{newtxtext,newtxmath}

\def\lsim{\;\raise0.3ex\hbox{$<$\kern-0.75em\raise-1.1ex\hbox{$\sim$}}\;}
\def\gsim{\;\raise0.3ex\hbox{$>$\kern-0.75em\raise-1.1ex\hbox{$\sim$}}\;}
\newcommand{\vect}[1]{\boldsymbol{#1}}
\newcommand{\kms}{\,km\,s$^{-1}$}
\newcommand{\gcm}{\,g\,cm$^{-3}$}
\newcommand{\ergs}{\,erg\,s$^{-1}$}
\newcommand{\Msun}{\,$M_{\odot}$}
\newcommand{\MsunYr}{\,$M_{\odot}$\,yr$^{-1}$}
\newcommand{\pccell}{\,pc\,cell$^{-1}$}

\newcommand{\textmc}[1]{\textsc{\scalefont{1.15}#1}}

\title[Inside the core of a young massive star cluster]{Inside the core of a young massive star cluster: 3D~MHD simulations}

\author[D. V. Badmaev, A. M. Bykov, and M. E. Kalyashova]{
D. V. Badmaev\thanks{E-mail: badmaev@astro.ioffe.ru},
A. M. Bykov,
and M. E. Kalyashova
\\
Ioffe Institute, 26 Politekhnicheskaya St, 194021 Saint-Petersburg, Russia
}

\date{Accepted XXX. Received YYY; in original form ZZZ}

\pubyear{2022}

\begin{document}
\label{firstpage}
\pagerange{\pageref{firstpage}--\pageref{lastpage}}
\maketitle

\begin{abstract}
Young massive star clusters inhabit regions of star formation and play an essential role in the galactic evolution. They are sources of both thermal and non-thermal radiation, and they are effective cosmic ray accelerators. We present the 3D magnetohydrodynamic (MHD) modeling of the plasma flows in a young compact cluster at the evolutionary stage comprising multiple interacting supersonic winds of massive OB and WR stars. The modeling allows studying the partitioning of the mechanical energy injected by the winds between the bulk motions, thermal heating and magnetic fields. Cluster-scale magnetic fields reaching the magnitudes of $\sim$ 300 $\mu$G show the filamentary structures spreading throughout the cluster core. The filaments with the high magnetic fields are produced by the Axford-Cranfill type effect in the downstream of the wind termination shocks, which is amplified by a compression of the fields with the hot plasma thermal pressure in the central part of the cluster core. The hot ($\sim$~a~few~keV) plasma is heated at the termination shocks of the stellar winds and compressed in the colliding postshock flows. We also discuss a possible role of the thermal conduction effects on the plasma flow, analyse temperature maps in the cluster core and the diffuse thermal X-ray emission spectra. The presence of high cluster-scale magnetic fields supports the possibility of high-energy cosmic ray acceleration in clusters at the given evolutionary stage.      
\end{abstract}

\begin{keywords}
galaxies: star clusters: general -- stars: winds, outflows -- stars: massive -- ISM: magnetic fields -- MHD
\end{keywords}



\section{Introduction}
\label{sec:intro}
A sizeable part of massive stars are born, evolve, and end their lives in clusters, which generally represent gravitationally bound groups of stars of the common origin \citep[][]{LL03,dW05,Par07,PZ10}. Of particular interest are young massive star clusters (YMSCs) since they are the dominant population in starburst galaxies and likely play a crucial role at the cosmic star-formation peak \citep[e.g.][]{Fig08,Dav10,Tan14,Ren18,Ada20}. There are arguably no better testing grounds for the stellar formation and evolution theories than the YMSCs, which can probably host diverse populations of coeval stars \citep[e.g.][]{Kra20}. Starburst galaxies are considered as likely sources of high-energy gamma-ray and neutrino emission \citep[e.g.][]{LW06,Tam14}, which may be due to YMSCs suggested as the promising sites of effective cosmic ray production \citep[e.g.][]{Byk14,Aha19,Byk20,Gupta20,Romero20,Morlino21,Vieu22,Bhadra22}. Moreover, the high-energy radiation and cosmic rays accelerated in YMSCs can provide a feedback effect which regulates star formation and disc structure in galaxies \citep[see e.g.][]{2021ApJ...910..126S}.

These clusters are observed in broad ranges of masses, ages, and densities \citep[][]{Pfl09,Kru19}. The most massive, compact and young of them ($M_{\mathrm{cl}}\gtrsim10^{4}$\Msun, $r_{\mathrm{cl}}\sim1$\,pc, $t_{\mathrm{cl}}\sim10^{6}$\,yr) are called 'super star clusters' (SSCs) and might be found in different types of galaxies with high star-forming rate \citep[][]{Whi00,Ada20}. The Local Group contains about a dozen well studied SSCs, such as NGC 3603 \citep[MW:][]{Dri95}, Westerlund 1/2 \citep[MW:][]{Cla05,Zei15}, Arches \citep[MW-GC:][]{Fig02}, Quintuplet \citep[MW-GC:][]{Fig99}, R136 \citep[LMC-30 Dor:][]{Mas98}. These SSCs contain dozens and even hundreds of prominently bright OB, Wolf-Rayet (WR), cool super- and hypergiant stars that are enclosed within the cluster cores of a parsec scale size.

Luminous massive stars possess powerful stellar winds ($P_{\mathrm{w}}\sim\,10^{36-38}$\ergs) \citep[e.g.][]{Cyr13}. Strong radiation generates a force that pushes out the upper layers of the stellar atmosphere initiating significant mass loss ($\dot{M}\sim10^{-4}-10^{-6}$\MsunYr) in the form of the wind \citep[][]{PPK86,KP00}. Being highly supersonic, the expanding wind material creates a bubble surrounded by a cool shell of the ISM gas that has been swept up by the forward shock front \citep[][]{Cas75,Wea77}. The geometry of the wind and its bubble depend on  stellar rotation \citep[e.g.][]{Ign96,Lan99}, structure and magnetisation of both wind \citep[e.g.][]{CL94,GS99,Mey21} and ISM \citep[e.g.][]{van15}. Typical radius of the solitary bubble is about $10-100$\,pc, which is much bigger than the mean distance between the stars in  SSCs. As a result, the interaction of multiple tightly packed supersonic winds leads to the formation of complex shocked environment inside the cluster core.

Due to the powerful kinetic energy release and its thermalisation in the cluster core \citep[see e.g.][]{Chv85,SH03} cluster scale wind can be produced. The structure and characteristics of this wind are regulated by the partitioning of the injected by the fast winds energy between the bulk motions, thermal and magnetic energies. \citet[][]{Can00} presented an analytic and first numerical model describing the 'cluster wind' flow that results from the multiple interaction of steady, mass-loaded stellar winds inside an SSC. The model was based on the pioneering work of \citet[][]{Chv85} who studied the problem of an adiabatic spherically symmetric wind driven by the supernova explosions in a starburst galaxy nucleus.

Modeling of the regions where winds interact is needed to estimate the characteristics and the structure of the hot plasma flows and magnetic fields in YMSCs. Three-dimensional (3D) hydrodynamic (HD) simulations of an SSC with interacting stellar winds were presented by \citet[][]{Rag01} and \citet[][]{RG07,RG08}. Particularly, \citet[][]{RG07} studied in much detail the effects of different stellar distributions within a cluster. The role of radiative energy losses and its imprint on the subsequent evolution of the SSCs was also thoroughly studied \citep[see e.g.][]{Sil04,Wun11,Ten07,Ten13}. 3D radiation-HD simulations were performed by \citet[][]{Wun17}, where they studied the impact of rapidly cooling winds on secondary stellar generations. In their 3D HD model \citet[][]{CR15} considered a supernova explosion in the context of massive stellar association, together with the effects of wind metallicity and thermal conduction. Detailed 3D HD modeling of massive star cluster feedback on the leftover from its parent molecular cloud was performed by \citet[][]{RP13,RP14}. \citet[][]{Bla20} discussed the impact of the wind induced turbulence on the subsequent star formation in the clusters. None of these studies considered magnetic field and its probable amplification rates inside the cluster cores, while the magnetic fields of different scales in the cluster play a crucial role in modeling of high-energy particle acceleration, non-thermal radiation from radio to gamma rays and neutrino emission.

Both thermal X-ray emission \citep[e.g.][]{Luo90,PP10} and non-thermal radiation \citep[e.g.][]{PRV21} are expected to be the signatures of the interactions of the fast winds in colliding wind binaries. The parsec scale size diffuse X-ray emission zones with luminosity $\sim10^{34}$\ergs can be produced by multiple stellar winds in compact clusters \citep[][]{Roc05}. X-ray emission of compact clusters is a subject of a number of observations.
For instance, Westerlund 1 was observed by \textit{Chandra} and \textit{XMM-Newton} and its features were analysed by \citet[][]{Mun06} and \citet[][]{KavanaghWd1}. While \textit{XMM-Newton} data point at the thermal nature of the emission,  \textit{Chandra} data allow the non-thermal interpretation of the spectrum.
Thermal and possibly non-thermal X-ray emission was detected by \textit{Chandra} in the Galactic YMSCs Arches and Quintuplet \citep[][]{Law04, Wan06}. 

Radio, infrared, and X-ray observations revealed a bunch of YMSCs in the Local Group. Evidence of high-energy particle acceleration in Doradus 30 region of LMC containing compact star clusters was found by \citet[][]{Lopez20} with \textit{NuSTAR} and \textit{XMM-Newton} observations. The R136 cluster in 30 Dor contains $>30$ hot and luminous O2-3 class stars, as well as $\sim$ 10 extremely bright WNh type Wolf-Rayet  stars \citep[][]{Crw10,Dor13}. Recent {\sl eROSITA} study of a bright extended X-ray source which coincides with R136 indicated a presence of a non-equilibrium
ionisation keV-temperature plasma emission and possibly a non-thermal component with a hard photon index \citep[][]{Sasaki22}. 

Physical interpretation of the available observational data and studying the impact of  YMSCs on the star-forming regions in galaxies as well as their possible role as high-energy particle accelerators and gamma-ray sources motivate the complex modeling of plasma flows and magnetic fields produced by powerful stellar winds in clusters.       

We present here the 3D MHD simulations of a SSC core with multiple interacting winds on the base of freely-distributed numerical code \textmc{pluto} \citep[][]{Mig07,Mig12,Mig18}. We examine the structure of the flows and provide the maps of the key characteristics of the gas inside the cluster. We focus on the magnetic field morphology and amplification inside the core, which is very important in the context of particle acceleration in YMSCs. Finally, we discuss the thermalisation of the wind energy, which is essential for the cluster-scale wind formation, and present the illustrative thermal spectra of a cluster core, exploring its features.

This paper is organized as follows. In $\S$~\ref{sec:numsim} we describe in detail the idea lying behind our model, numerical scheme, some microphysics details, initial and boundary conditions of the SSC core simulation. We deliver the results of the simulations in $\S$~\ref{sec:res}. In $\S$~\ref{sec:dis} we discuss the magnetic field, thermalisation and thermal emission in the cluster core. We briefly conclude our findings in $\S$~\ref{sec:con}.

\section{Numerical simulations}
\label{sec:numsim}

\begin{table*}
	\centering
	\caption{Nomenclature and grid parameters of the models.}
	\label{tab:tab1}
	\begin{tabular}{lccccccccr} 
		\hline
		Model & $R_{\mathrm{cl}}$\,(pc) & $\dot{E}_{\mathrm{cl}}$\,($10^{39}$\ergs) & $\dot{M}_{\mathrm{cl}}$\,($10^{-3}$\MsunYr) & $\Delta$\,($10^{-3}$\pccell) &
		{Included physics}\\
		\hline
		500:Base-UpScale & 2.0 & 1.15 & 2.23 & 8 & MHD+Cool\\
		250:Base & 2.0 & 1.15 & 2.23 & 16 & MHD+Cool\\
		125:Base & 2.0 & 1.15 & 2.23 & 32 & MHD+Cool\\
		125:Base-TC & 2.0 & 1.15 & 2.23 & 32 & MHD+Cool+TC\\
	    \hline
		125:Low-Mdot & 2.0 & 1.15 & 1.71 & 32 & MHD+Cool\\
		125:Low-MEdot & 2.0 & 0.57 & 1.63 & 32 & MHD+Cool\\
		\hline
		125:Low-Mdot-S & 1.0 & 1.15 & 1.71 & 16 & MHD+Cool\\
        125:Low-MEdot-S & 1.0 & 0.57 & 1.63 & 16 & MHD+Cool\\
		\hline
	\end{tabular}
\end{table*}

\begin{figure}
	\includegraphics[width=8.75cm]{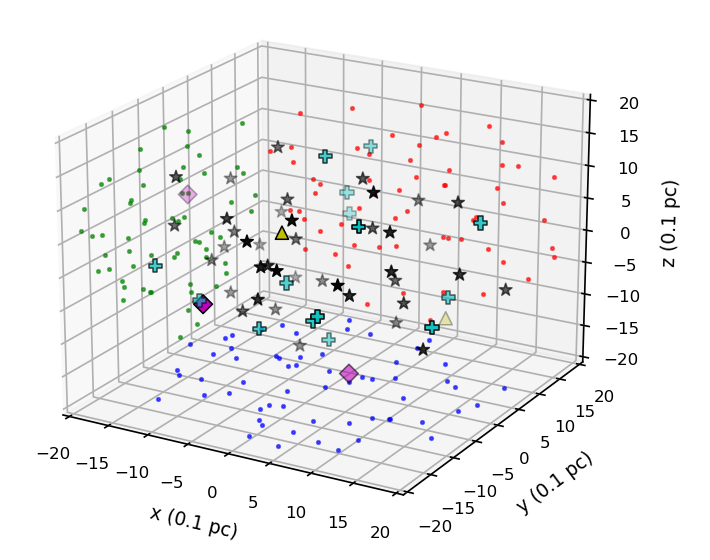}
	\vspace*{-5mm}
    \caption{Stellar distribution inside the cluster. Black stars stand for O-type stars, cyan crosses - Wolf-Rayet stars, purple diamonds - YSG stars, yellow triangles - RSG stars. Opacity of the markers shows the depth relatively to the view point. Red, green, and blue dots represent $xz$-, $yz$-, $xy$-plane projections respectively.}
    \label{fig:disrtib}
\end{figure}

\subsection{Building of the cluster}
\label{sec:spop}

To build a realistic model of a SSC one needs to pick up an object with a well known stellar census as a reference. There are only a few SSCs in the Galaxy that have been resolved up to a point-source scale. One of the most interesting of SSCs, Westerlund 1 (Wd1), located in the relative proximity to the Solar system \citep[$d\approx4.2$\,kpc;][]{Neg22}, was extensively studied by the observatories with high angular resolution in radio \citep[][]{Fen18,And19}, optical \citep[][]{Cla05,Neg10} and X-ray bands \citep[][]{Mun06,Cla08}. These observations revealed rich populations of OB, WR, and even rare cool supergiant stars (CSG) confined in the cluster core of a few parsec scale size. In comparison with the other known SSCs of the Local Group Wd1 turns out to be one of the most massive, yet given the poor sample of these known SSCs \citep[about a dozen of them;][]{PZ10} it is hard to draw any conclusions about Wd1's peculiarity. Probably, in the galaxies with higher star-forming rates such massive SSCs are more common. Besides its importance in general problems of star formation and evolution of SSCs, the cluster was recently recognized as a powerful source of non-thermal radiation \citep[][]{Aha19,Wd1_HESS22}. Based on these facts we chose Wd1 as the most opportune reference stellar system for our simulations.   

In this work we focus our attention on two groups of stars. The first one represent the late O/WR stars with fast and energetic winds ($v_{\mathrm{\infty}}\sim1000$\kms) that should dominate the overall dynamics, and the second one represent the CSG stars that are sources of severe mass loss ($\dot{M}\sim10^{-4}$\MsunYr) in the cluster. In particular, the well-established observations of Wd1 revealed about a 100 of prominently bright OB-type sources \citep[][]{Neg10,Neg22}, 24 WR \citep[all late WC/WN types;][]{Cro06}, 6 YSG, and 4 RSG stars \citep[][]{Cla05}. The latest analysis of Gaia EDR3 data \citep[][]{Neg22} revealed an elliptical shape of Wd1 with a minor axis of the ellipse estimated to be 3.7 pc, yet the whole population of CSGs and WRs is enclosed inside a central sphere of 2 pc in radius \citep[][]{Bea21}. We do not pretend here to mimic, in any way, neither the observed stellar distribution patterns nor the full scale cluster. In this way, it was enough for us to consider a uniform random distribution of O/WR and CSG populations inside a sphere of 2 pc in radius. Certainly another spatial distribution of stars would provide somewhat different flow morphology in detail. However, we believe that the most important features of this modeling concerned energy partition and magnetic field amplification should hold and scale with the kinetic power of the most massive stars.

The star cluster simulation has a set of crucial parameters: mechanical energy input ($\dot{E}_{\mathrm{cl}}$), mass loading ($\dot{M}_{\mathrm{cl}}$), compactness, and stellar population (in terms of evolutionary stages). The first two parameters have a vast space for variation ($\sim$ a decade wide) due to empirical uncertainties. In our modeling we keep the observed proportion between different stellar types, yet roughly halving the total cluster population, thus, mimicking Wd1 on a lower scale in terms of number of stars. This implies: 40 O-type, 15 WR, 3 YSG, and 2 RSG stars. It is technically challenging to do a full scale model, because the higher number of stars requires higher grid resolution and, therefore, dramatically raises the computational cost. To compensate roughly half the number of stars of each type we double their mass loss rates (as the most uncertain parameter) in the 'base' simulation. All the other stellar wind parameters are kept constant throughout the simulations (see Table~\ref{tab:tab2}). As a result, we keep $\dot{E}_{\mathrm{cl}}$ and $\dot{M}_{\mathrm{cl}}$ at the levels estimated for Wd1 on the base of observational and \citet[][]{Eks12} evolutionary data: $\dot{E}_{\mathrm{cl}}\sim10^{39}$\ergs, $\dot{M}_{\mathrm{cl}}\sim2\times10^{-3}$\MsunYr. The compactness is determined here by the number of stars enclosed inside the cluster radius, $r_{\mathrm{cl}}=2$\,pc (1 pc for the models with 'S' prefix, see Table~\ref{tab:tab1}).

For the stellar sampling inside the computational domain we use a simple technique described in detail by \citet[][]{RG08}. Generally, this technique allows random stellar sampling with variable power-laws of the central radius. In our simulations we decided to keep a uniform stellar distribution inside the system mainly because of the idea to consider only the central bulge of the cluster, where the stellar density varies negligibly. We also improved the algorithm of sampling to prevent any overlapping of stellar positions up to a small area around each star, i.e. the distance between any pair of stars in the cluster is not shorter than the minimal allowed distance. This restriction is related to the computational constraint on the grid resolution. For the resulting stellar distribution see Fig.~\ref{fig:disrtib}.

Eventually, we performed 1 base (250:Base) simulation and 3 simulations investigating the effects of thermal conduction (125:Base-TC) and resolution (125/250/500:Base) on the magnetic field, flows geometry and thermal emission (see Table~\ref{tab:tab1}). The 'UpScale' prefix means that we used the final results from 250:Base simulation and extended it on a few dynamical times forward using the grid of $500^{3}$. We carried out 4 additional simulations to investigate the dependence of the cluster's thermal spectrum behaviour on the energy and mass loss tweaks (125:Low-M/MEdot) and length scale (125:Low-M/MEdot-S), for the details see $\S$~\ref{sec:them}. The 'Low-Mdot' and 'Low-MEdot' prefixes stand for the simulations where the mass loss rates of either CSG or O/WR-type stars do agree with estimations of \citet[][]{Fen18}, respectively, without being artificially enhanced.

\subsection{Governing equations}
\label{sec:goveq}

The simulations were performed using the well-proven open source code \textmc{pluto} \citep[][]{Mig07,Mig12,Mig18} based on the Godunov method, and created specifically for problems of computational astrophysics. According to our problem, the code integrates the following set of non-ideal magnetohydrodynamic equations:
\begin{gather}
    \frac{\partial\rho}{\partial{t}}+\vect{\nabla}\cdot\left(\rho\vect{u}\right)=0,\label{1}\\
    \frac{\partial\vect{m}}{\partial{t}}+\vect{\nabla}\cdot\left(\vect{mu}-\vect{BB}+\vect{I}p_{\mathrm{tot}}\right)=0,\label{2}\\
    \frac{\partial{E}}{\partial{t}}+\vect{\nabla}\cdot\left[\left(E+p_{\mathrm{tot}}\right)\vect{u}-\vect{B}\left(\vect{u}\cdot\vect{B}\right)\right]=\vect{\nabla}\cdot\vect{F}_{\mathrm{c}}+\Phi\left(T,\rho\right),\label{3}\\
    \frac{\partial\vect{B}}{\partial{t}}+\vect{\nabla}\cdot\left(\vect{uB}-\vect{Bu}\right)=0,\label{4}
\end{gather}
where $\vect{m}=\rho\vect{u}$ represents the momentum density vector of a control volume, $\vect{B}$ is the magnetic field vector, $\vect{I}$ is the identity matrix, $p_{\mathrm{tot}}=p+\vect{B}\cdot\vect{B}/2$ is the total pressure. The total energy density of the systems reads,
\begin{equation}
    E=\frac{p}{\gamma-1}+\frac{\vect{m}\cdot\vect{m}}{2\rho}+\frac{\vect{B}\cdot\vect{B}}{2},
\end{equation}
and the sound speed, $c_{\mathrm{s}}=\sqrt{\gamma{p}/\rho}$, closes the above system of equations, where $\gamma=5/3$ is the adiabatic index. The source term $\Phi\left(T,\rho\right)$ on the right-hand side of the total energy conservation equation represents optically thin radiative losses and heating. The plasma heat flux is determined by the vector $\vect{F}_{\mathrm{c}}$.  

We took into account the gains and losses by optically thin radiative processes following the recipe from \citet[][]{Mey14,Gre19} for the case of photo-ionization equilibrium (PIE):
\begin{equation}
    \Phi\left(T,\rho\right)=n_{\mathrm{H}}\Gamma\left(T\right)-n_{\mathrm{H}}^2\Lambda\left(T\right),
\end{equation}
where $\Gamma\left(T\right)$ and $\Lambda\left(T\right)$ are the radiative heating and cooling rates, respectively, and $n_{\mathrm{H}}$ is the hydrogen number density. 

This system of equations is solved by use of the unsplit second-order Runge-Kutta algorithm with linear reconstruction of the variables between adjacent cells together with the combination of HLLD \citep[][]{MK05,Mig07} and HLL \citep[][]{Har83} Riemann solvers. The divergence-free condition for the magnetic field is ensured by the Hyperbolic Divergence Cleaning algorithm \citep[][]{Ded02} in the whole computational domain. The time-marching algorithm is controlled by the standard Courant-Friedrichs-Lewy parameter that we set to $C_{\mathrm{cfl}}=0.2$.

\begin{table}
	\centering
	\caption{Stellar wind parameters.}
	\label{tab:tab2}
	\begin{tabular}{lccr}
		\hline
		Type & $v_{\mathrm{\infty}}$\,(\kms) & $\dot{M}$\,(\MsunYr) & $\dot{E}$\,(\ergs)\\
        \hline
         O-wind & 2300 & $5.35\times10^{-6}$ & $9.01\times10^{36}$\\
		 WR-wind & 1600 & $6.50\times10^{-5}$ & $5.27\times10^{37}$\\
		 YSG-wind & 50 & $2.30\times10^{-4}$ & $1.83\times10^{35}$\\
		 RSG-wind & 35 & $1.75\times10^{-4}$ & $6.82\times10^{34}$\\
		\hline
	\end{tabular}
\end{table}

\subsection{Electron equilibration}
\label{sec:eleq}
The effects of plasma cooling and thermal conduction are sensitive to electron temperature, as well as radiation spectra. Since the single-fluid MHD models cannot directly provide information on the electron and ion temperatures separately, they need to be supplemented with the appropriate recipe to estimate the local electron temperatures. The main source of the mechanical power in  stellar clusters are  supersonic winds of young massive stars and supernovae. Therefore, collisionless shock waves provide  bulk plasma heating in clusters. Electron heating efficiency at collisionless shocks and the thermal energy partitioning between electrons and ions depend on the Mach number and magnetisation of the flow. While these issues are not well understood yet, some differences in electron heating  between the strong and moderate shocks are established \citep[see e.g.][]{Ghavam_13,vink15,el_shock20}.

In the kinetic particle-in-cell (PiC) models of electron heating by strong transverse shocks (with $m_{\mathrm{i}}/m_{\mathrm{e}}$ ratio 200--400) made by \citet[][]{el_shock20} a superadiabatic electron heating was demonstrated. The downstream electron temperatures for the shocks with Alfvenic Mach numbers $M_{\mathrm{A}}$ between 20 and 70  were within a range of 0.11--0.19 $T_{\mathrm{p}}$. Also, \citet[][]{el_shock20Tran} performed 2D PiC simulations of collisionless shocks with the magnetosonic Mach numbers in the range of 1–10 and the upstream plasma magnetisation parameter $\beta=8\pi n T/B^2 <0.4$. They found that post-shock electron-to-ion temperature ratio decreases from $\sim$ 1 to 0.1 with the increase of the magnetosonic Mach number. 

The electrons in the downstream  of a strong shock heated up to $\sim$ 0.1 $T_{\mathrm{p}}$ can reach plasma equilibrium temperatures due to Coulomb collisions \citep[e.g.][]{BPP08}, though some extra collisionless heating cannot be excluded as well. Indeed, plasma, shocked by interacting winds, is most likely highly turbulent over a broad range of spatial scales down to the kinetic scales, which cannot be resolved by the MHD codes. The ratio of ion-to-electron heating due to the dissipation of the cascading turbulence in the kinetic scales  depends on the plasma magnetisation parameter as it was demonstrated, e.g., by \citet[][]{Howes10} in the case of the Alfvenic turbulence. In the solar wind the electron temperature is anisotropic and the electron heat fluxes, measured by spacecrafts, show the heat flux regulation by the whistler mode waves \citep[see e.g.][and the references therein]{Art19}. Namely, the ratio of the electron heat to the free-streaming heat flux is inversely proportional to $\beta_e = 8\pi n T_{e{\parallel}}/B^2$. 

The accurate treatment of the electron kinetics in the global modeling of the plasma flows and mapping of the electron temperatures in the inner parts of the YMSC are unfeasible and beyond the scope of this paper. We used here some simplified recipes based on the current knowledge to estimate the electron temperatures from the single-fluid MHD \textmc{pluto} simulations.

\subsubsection{Thermal conduction}
\label{sec:thcon2}
The heat conduction effects arise mostly due to the electron transport. As discussed above, the kinetics of electron heating in the hot rarefied plasma with the weak Coulomb collisions is a non-trivial task.  Nevertheless, it is feasible to estimate the possible effect of the heat transport on the structure of the MHD plasma flows by considering two limiting cases. One model assumes a relatively inefficient electron heating and neglects the thermal conduction. Another model assumes some efficient non-adiabatic electron heating and the presence of the electron thermal conduction along the local magnetic field. To illustrate the limits of the electron heating/thermal conduction effects  we  rather speculatively assumed the electron temperature being close to that of ions. This may, in principle, be established by the processes like the collisionless plasma heating by turbulence at the kinetic scales which are beyond the ideally conductive single-fluid MHD. 
Indeed, the two-fluid MHD solar wind models which allow a comparison with the direct measurements revealed that the MHD turbulence dissipation is likely dividing the energy between protons and electrons with $\sim$ 60\% going to proton heating and $\sim$ 40\% into electron heating \citep[see e.g.][]{Breech09}.  
The comparison of the two limiting cases in our single-fluid 3D MHD simulations, presented in $\S$~\ref{sec:tc}, allows the rough estimate of the significance of the electron thermal conduction effects while study of kinetic effects is needed to address the issue properly.       

The standard approach to account for the anisotropic heat conduction is described in \citet[][]{Bra65} as:
\begin{equation}
    \vect{F}_{\mathrm{cl}}=\kappa_{||}\vect{b}\left(\vect{b}\cdot\nabla{T_{\mathrm{e}}}\right)+\kappa_{\perp}\left(\nabla{T_{\mathrm{e}}}-\vect{b}\cdot\nabla{T_{\mathrm{e}}}\right),
\end{equation}
where $\vect{b}=\vect{B}/|\vect{B}|$ is the magnetic field unit vector. The $\kappa_{||}$ and $\kappa_{\perp}$ are the conductive coefficients along and across the local magnetic field lines in collisional plasma, respectively. The unsaturated heat transfer along the magnetic field is given by \citep[][]{SH53,Spi62}:
\begin{gather}
\kappa_{||}=5.6\times{10}^{-7}{T_{\mathrm{e}}}^{5/2},\\
\kappa_{\perp}=3.3\times{10}^{-16}\frac{n_{\mathrm{H}}^2}{T_{\mathrm{e}}^{1/2}{|\vect{B}|}^2}.
\end{gather}
and we accounted for the transition to the saturated heat conduction regime \citep[][]{Hollweg76,Cow77,Bal13}. A smooth transition between the standard classical and the saturated heat fluxes can be approximated by the following expression \citep[e.g.][]{Orl08}:
\begin{equation}
    \vect{F}_{\mathrm{c}}=\frac{F_{\mathrm{sat}}}{F_{\mathrm{sat}}+|\vect{F}_{\mathrm{cl}}|}\vect{F}_{\mathrm{cl}},
\end{equation}
where $F_{\mathrm{sat}}$ is the saturated heat flux, that was calculated according to the analysis of the solar wind by \citet[][]{Bal13}, where they found a collisionless saturation of the heat flux at the level of $F_{\mathrm{sat}}\sim0.3{F}_{\mathrm{fs}}$, where ${F}_{\mathrm{fs}}$ stands for the so-called 'free streaming' heat flux:
\begin{equation}
    F_{\mathrm{fs}}=n_{\mathrm{e}}k_{\mathrm{B}}T_{\mathrm{e}}\sqrt{\frac{2k_{\mathrm{B}}T_{\mathrm{e}}}{\pi{m}_{\mathrm{e}}}}.
\end{equation}

\subsection{Initial and boundary conditions}
\label{sec:init}

We use a latitude-dependent stellar wind model based on the theory of \citet[][]{BC93} in the limit of large distances. The internal stellar wind boundaries follow the equations $(3)-(5)$ from \citet[][]{Lan99}, yet the terminal wind speed is now a function of the effective stellar temperature:
\begin{gather}
    v_{r}\left(\theta\right)=v_{\mathrm{\infty}}\left(1-\Omega\sin\theta\right)^{\gamma},\\
    v_{\mathrm{\infty}}=\zeta\left(T\right)v_{\mathrm{esc}}=\zeta\left(T\right)\sqrt{\frac{2GM\left(1-\Gamma_{\mathrm{Edd}}\right)}{R}},
\end{gather}
where $\zeta=2.6$ for O/WR and $\zeta=0.5$ for CSG wind \citep[cf.][]{KP00}, $\Omega=\sqrt{2}v_{\mathrm{rot}}/v_{\mathrm{esc}}$ is the stellar rotation parameter, and $\theta$ is the polar angle. The wind rotation is included by considering the following azimuthal velocity component  
\begin{equation}
    v_{\phi}\left(r,\theta\right)=v_{\mathrm{rot}}\frac{R}{r}\sin\theta,
\end{equation}
where $v_{\mathrm{rot}}=R\omega$ is the star's equatorial rotation velocity. The stellar mass $M$, effective temperature $T$, angular speed $\omega$, and Eddington gamma $\Gamma_{\mathrm{Edd}}=L/L_{\mathrm{Edd}}$ where derived from the evolutionary data of the \textit{Geneva} code group \citep[see][]{Eks12}. The mass loss rates $\dot{M}$ for O/WR and CSG winds were chosen so that to comply both with the \textit{Geneva} code data and, roughly, with the empirical estimations of \citet[][]{Fen18} and \citet[][]{And19} considering the clumping factor of unity (see Table~\ref{tab:tab2}).

The magnetic field structure in the freely expanding wind is treated as a Parker spiral \citep[see e.g.][]{Par58,CL94,GS99,Mey21}:
\begin{gather}
    B_{r}\left(r\right)=B_{\mathrm{s}}\left(\frac{R}{r}\right)^{2},\\
    B_{\phi}\left(r\right)=B_{\mathrm{s}}\left(\frac{R}{r}\right)\left[\frac{v_{\phi}\left(r,\theta\right)}{v_{r}\left(\theta\right)}\right]\left(\frac{r}{R}-1\right),
\end{gather}
where $B_{\mathrm{s}}$ is a magnetic field strength at the stellar surface. This field has a split-monopole structure, which implies a sign flip when crossing $\theta=\pi/2$. Throughout all simulations we set $B_{\mathrm{s}}=100$\,G for O/WR type stars in the cluster \citep[e.g.][]{Sch17}.

The \textmc{pluto} code solves the equations $(\ref{1}-\ref{4})$ in 3D Cartesian coordinates $(x,y,z)$. The computational domain was extended in the intervals of $[-2;2]$ or $[-1;1]$\,pc ('S' models) in all directions. In our simulations we used uniform grids of $125^{3}$ ('125:' models), $250^{3}$ ('250:' model), and $500^{3}$ ('500:' model) cells that cover the domain (see Table~\ref{tab:tab1}). At the domain borders we used a modified 'free outflow' boundary condition that also prohibits any possible 'backflow' of the gas. Firstly, in the entire domain we initialized a magnetised ISM of constant temperature, density, and magnetisation: $T=8.5\times10^{3}$\,K, $n= 0.5$\,cm$^{-3}$, $\boldsymbol{B}=B\boldsymbol{e}_{y}$, where $B=3.5$\,$\mu$G \citep[e.g.][]{Mey17}. Then, we injected the winds of 60 stars as the inner boundary conditions inside small spherical volumes of 5 cells in radius that are randomly distributed in the domain (see Fig.~\ref{fig:disrtib}). The simulations were stopped at the integration time $t_{\mathrm{int}}=10^{4}$ yr, which is well after the establishing of a quasi-stationary flow regime.

\begin{figure}
	\includegraphics[width=8.7cm]{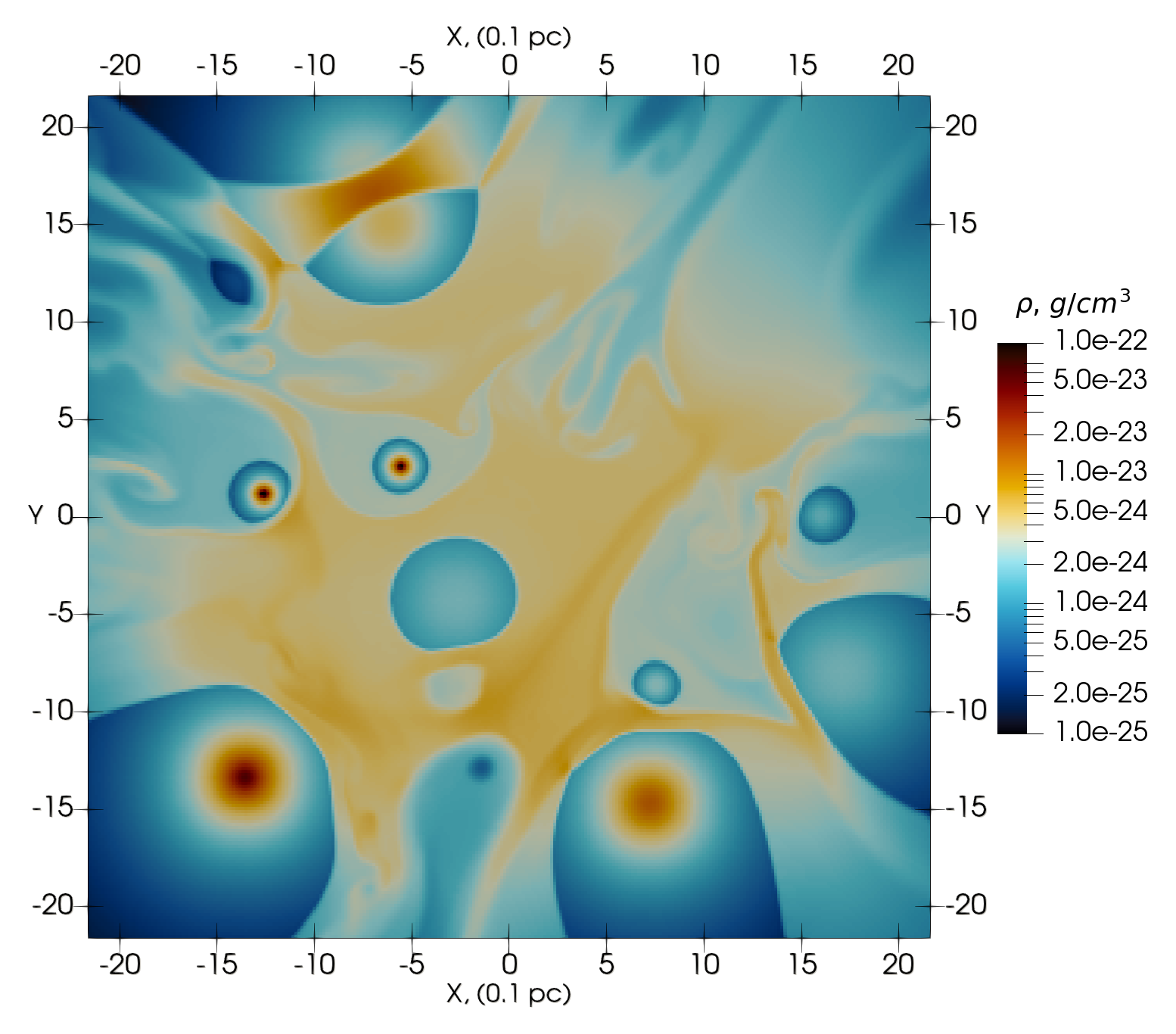}
	\vspace*{-5mm}
    \caption{Density map of the central $Oxy$-plane  of the cluster simulated with the single-fluid 3D MHD model.}
    \label{fig:dens}
\end{figure}

\begin{figure}
	\includegraphics[width=8.7cm]{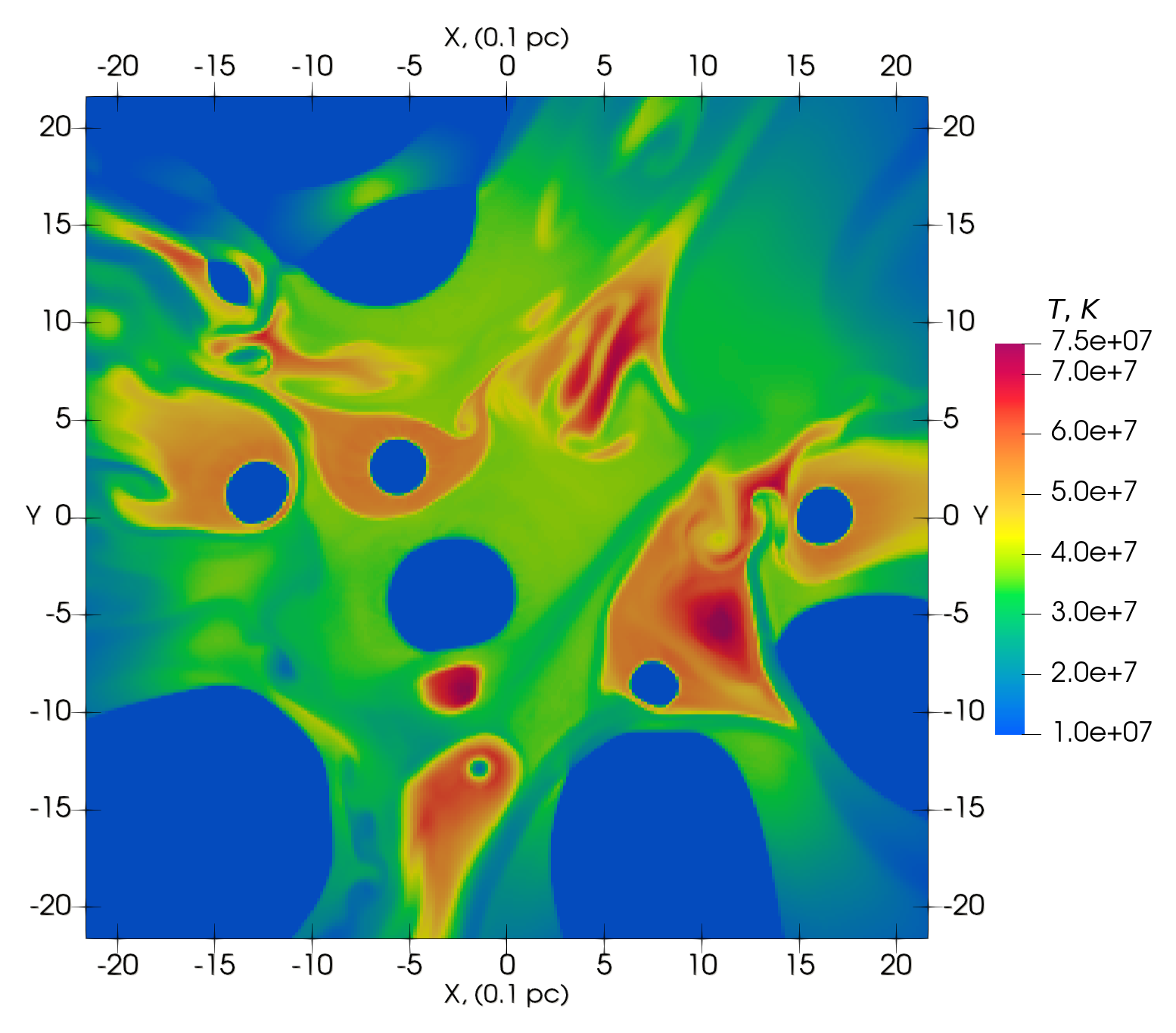}
	\vspace*{-5mm}
    \caption{Temperature map of the central $Oxy$-plane of the cluster simulated with the single-fluid 3D MHD model.}
    \label{fig:temp}
\end{figure}

\begin{figure}
	\includegraphics[width=8.7cm]{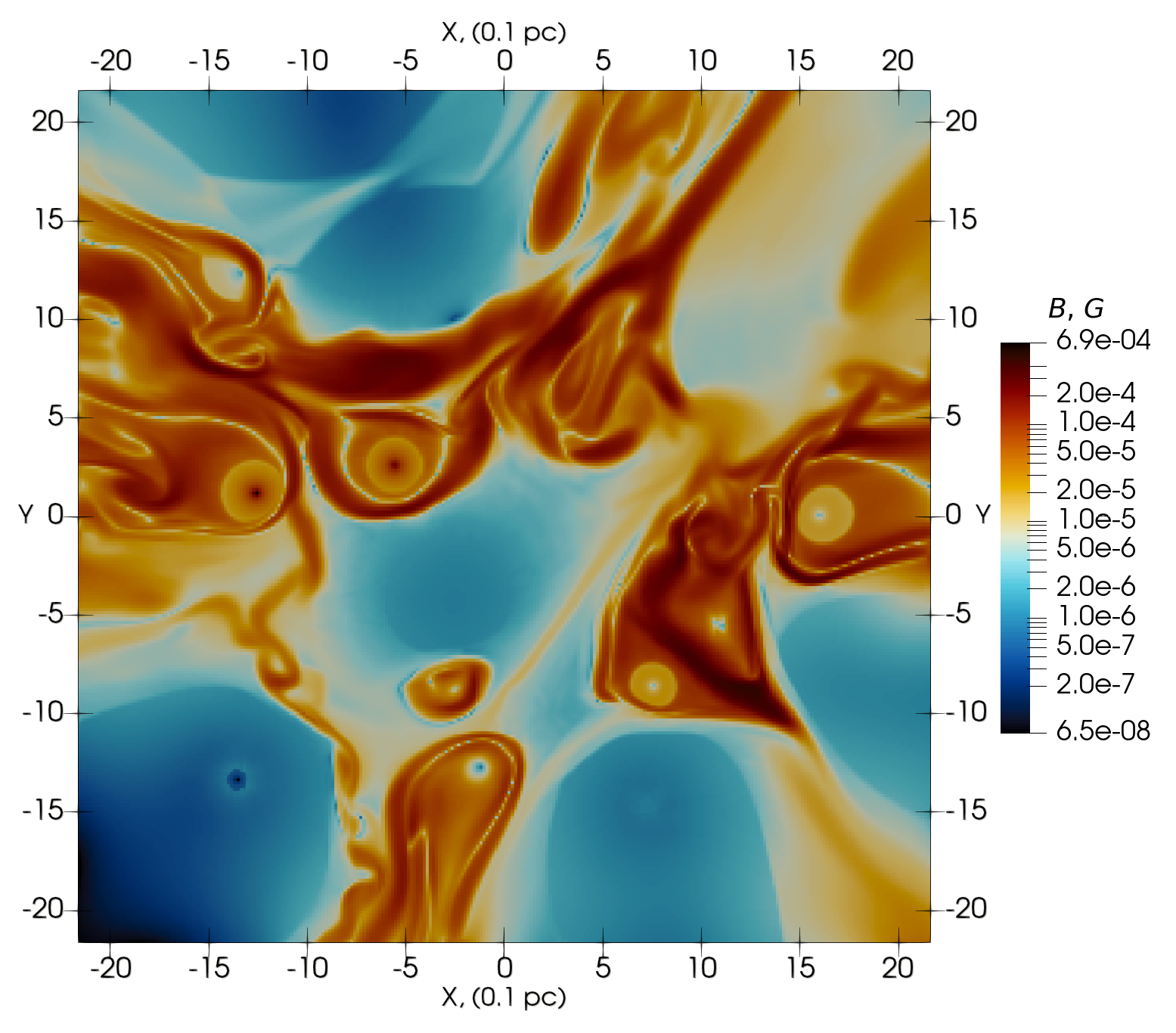}
	\vspace*{-5mm}
    \caption{Magnetic field map of the central $Oxy$-plane of the cluster simulated with the single-fluid 3D MHD model.}
    \label{fig:MF1}
\end{figure}

\begin{figure}
	\includegraphics[width=8.5cm]{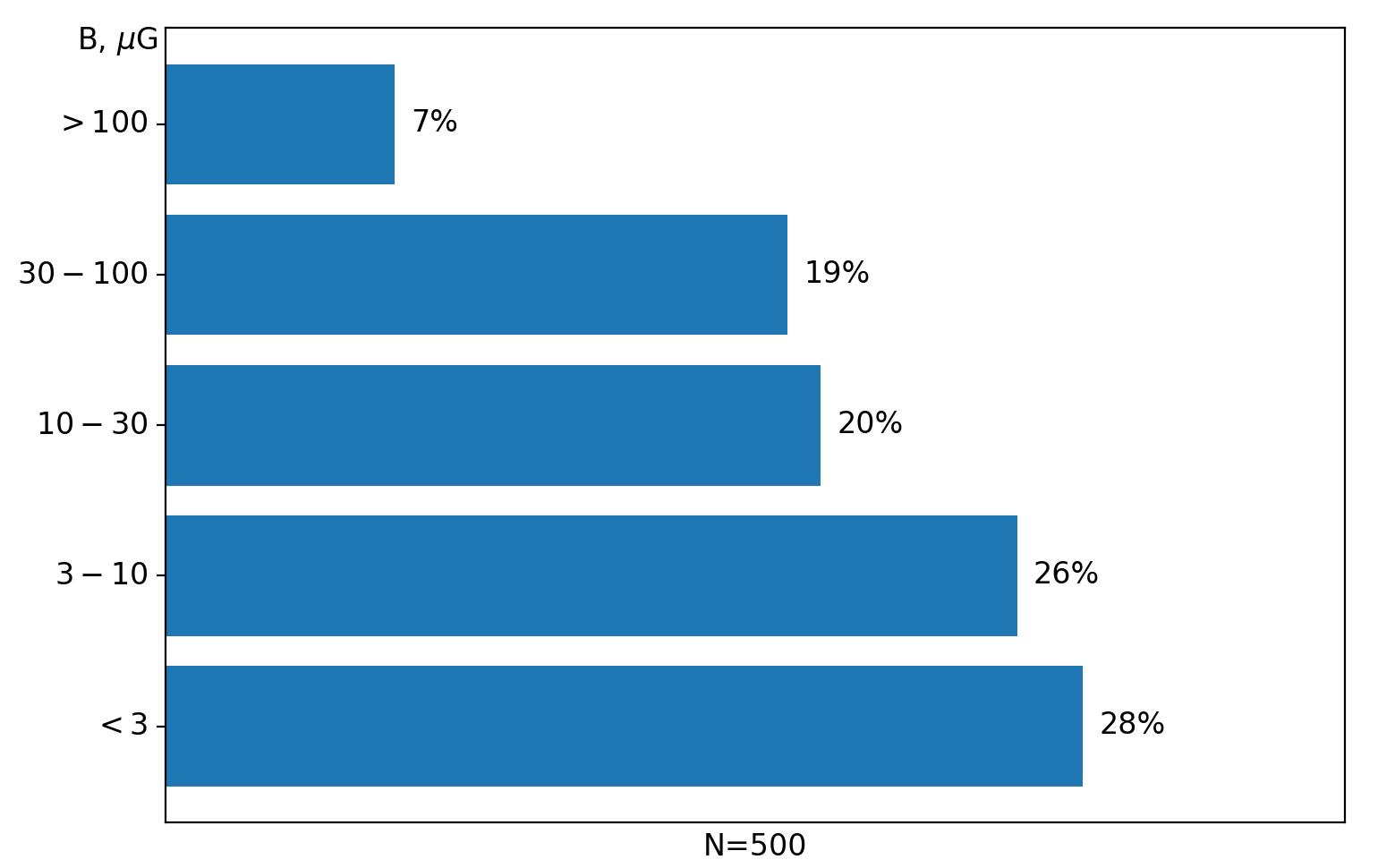}
	\vspace*{-1mm}
    \caption{The volume filling factors of the magnetic fields of different magnitudes in the simulated cluster (see Fig.~\ref{fig:MF1}). The factors are derived from the most refined ’500:Base’ simulation.}
    \label{fig:filfac}
\end{figure}

\begin{figure}
	\includegraphics[width=8.7cm]{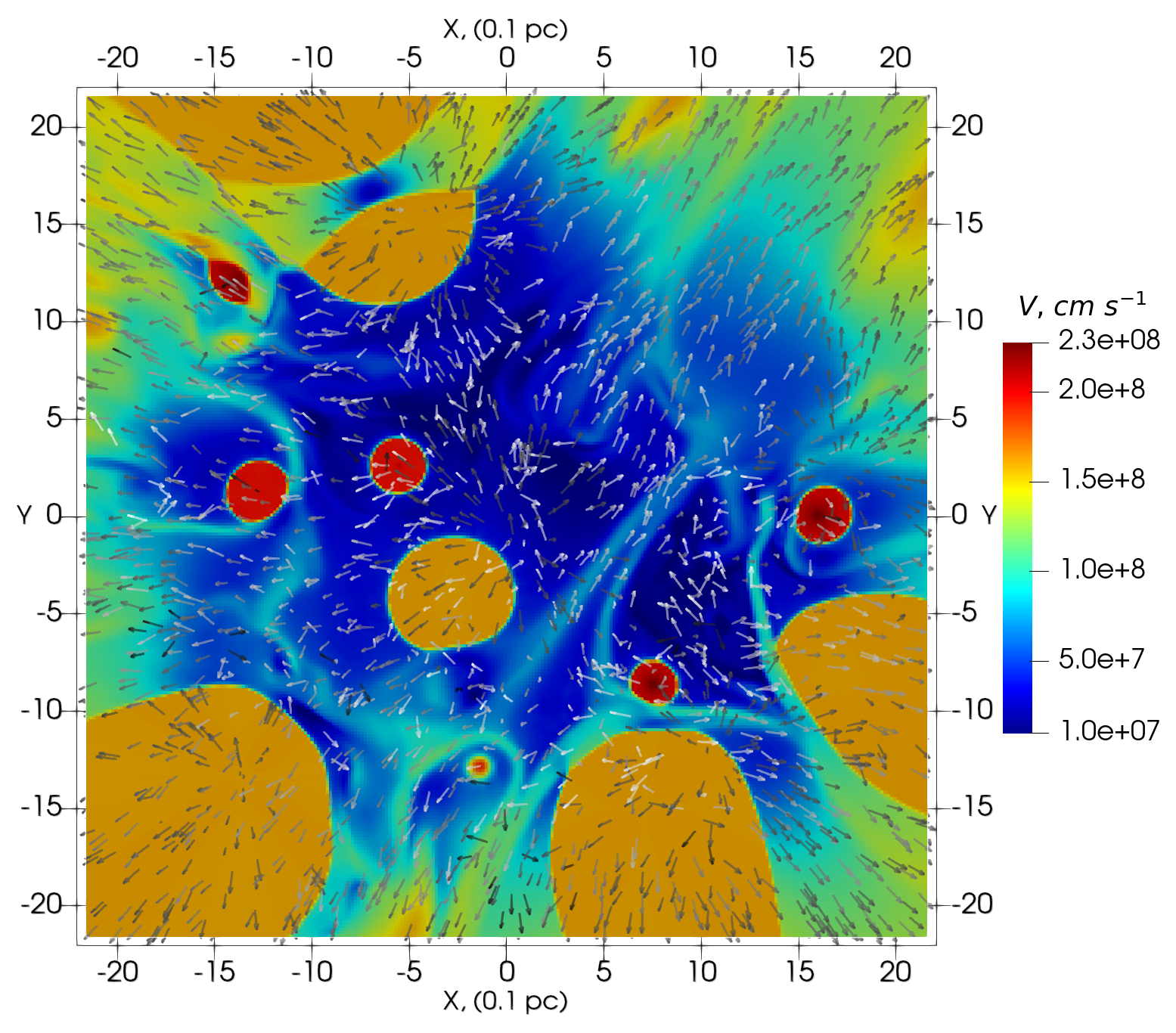}
	\vspace*{-5mm}
    \caption{Plasma bulk velocity field map of the central $Oxy$-plane of the cluster simulated with the single-fluid 3D MHD model. The arrows represent the direction of the local velocity field.}
    \label{fig:velo}
\end{figure}

\section{Results}
\label{sec:res}

\subsection{Structure of the flow and gas heating in the inner cluster}
\label{sec:flow}
Here we present the cluster core maps of the YMSC simulated with the 3D MHD model described above. Since we did not introduce any specific spatial symmetries in the stellar distribution (it is uniformly random) there is no any qualitative difference between choosing various central plane maps of the simulation variables to make illustrations. Thus, we made mostly the central maps normal to $z$ axis ($Oxy$-plane), as the other ones would show a similar general behavior of the flows, heating, and magnetic field amplification. Therefore, Figs.~\ref{fig:dens}-\ref{fig:velo} are the $Oxy$-plane maps of density, temperature, magnetic filed, and velocity magnitudes, respectively. These maps are given for '250:Base' simulation and reveal a complex geometry of flows.

In the Fig.~\ref{fig:velo} there is a bunch of prominent deformed circular regions of the supersonic winds that are limited by termination shocks and not yet thermalised: the red regions -- O-type winds, the orange ones -- WR winds. In fact, we have in the domain 60 sources of kinetic ram pressure that also carry almost a negligible amount of magnetic and thermal energy. Comparing all the Figures one could clearly see how the kinetic energy effectively transforms to the thermal and magnetic components. Typical flow speed at the central part of the domain, where most winds undergo head-on collisions, is about a few $100$\kms while the temperature there raises up to an X-ray level $\sim6\times10^{7}$\,K (see Fig.~\ref{fig:temp}) in the proximity of the O-type winds. It is also clearly seen that the central thermalised material accelerated by the thermal pressure gradient channels through the gaps between wind dominated regions at the periphery. The average gas density in the cluster is estimated $\sim1.1\times10^{-23}$\gcm (see Fig.~\ref{fig:dens}), while the total mass of the confined gas is $\sim{9.6}$\Msun. 

\begin{figure*}
	\includegraphics[scale=0.2973]{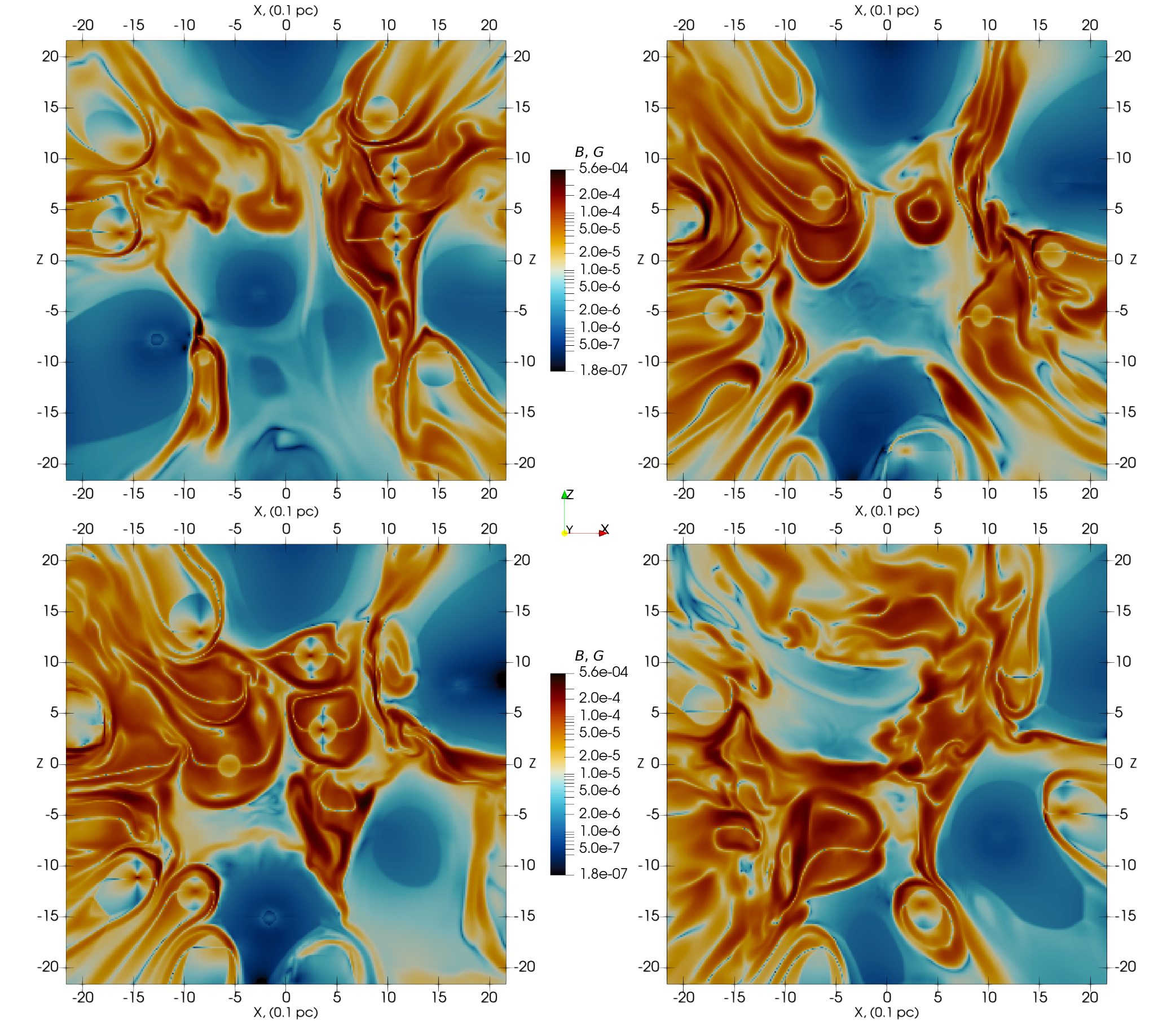}
	\vspace*{-1mm}
    \caption{Magnetic field magnitude maps captured in a series $xz$-planes, normal to the Fig.~\ref{fig:MF1}. The upper left corner: $y=-1.45$\,pc. The upper right corner: $y=-2.10$\,pc. The lower left corner: $y=2.35$\,pc. The lower right corner: $y=2.85$\,pc.}
    \label{fig:mfperp}
\end{figure*}

\begin{figure*}
	\includegraphics[scale=0.3]{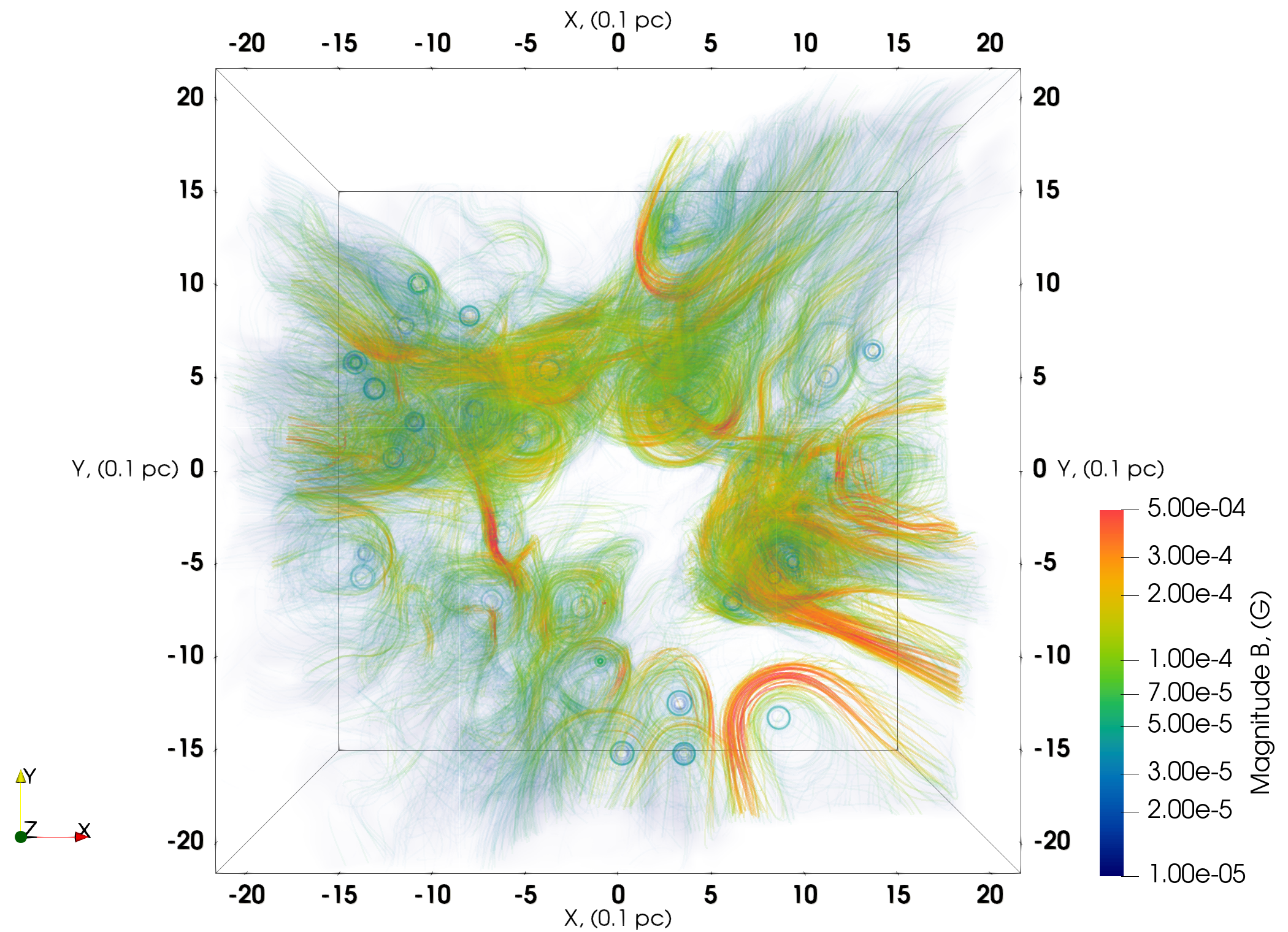}
	\vspace*{-1mm}
    \caption{Structure of the cluster magnetic fields rendering the filaments and bow-type shells downstream of the wind termination shocks simulated with 3D MHD model. Blue circles are the positions of the stars producing the fast winds carrying magnetic field.}
    \label{fig:3dmf}
\end{figure*}

In the regions of high compression, especially near the central strongly suppressed O-type sources, the magnetic field is amplified up to $\sim$ a few 100\,$\mu$G, as the magnetic flux conserves. Indeed, the highest magnitudes of the field are observed in the vicinity of O-type stars, where the expelled gas is swept up to and strongly compressed by the dominating WR-type winds (see the extensive blue 'voids'). In the Figs.~\ref{fig:MF1},~\ref{fig:mfperp} one could clearly see the filamentary structure of compressed magnetic field. The random uniform distribution of the stars in the domain ensures that the wind-wind collisions take place in various cases of mutual arrangement of pairs of stars. In the Fig.~\ref{fig:3dmf} we present a 3D render of the magnetic field structure, where its seen how the bow-shock structures are correlated with the amplified magnetic field regions. 

As expected, the CSG winds do not contribute notably to the overall dynamics of the resulting thermalised flow, but they turn out to be sufficiently inert at the time-scale of $\sim$ a few $10^{4}$\,yr, as their wind blown bubbles surrounded by dense ($\rho\gtrsim10^{-21}$\gcm) and thin shells keep their shape against the cluster core medium, see Fig.~\ref{fig:thermal}. Only the ones placed near the domain edges, where the cluster wind accelerates to $\sim1000$\kms, did form a bow shock structure. Still, the CSG population may change the thermal spectrum of the cluster introducing sizeable volumes filled with dense and cold ($T\lesssim10^{4}$\,K) material to the cluster core medium.

We also studied how the resolution affects on the low scale structure of the flows and magnetic fields. As the resolution increases in '125:Base', '250:Base', and '500:Base' simulations, we found out that the general shape of flows and quantitative data of all variables do not change notably, except the low scale magnetic field features \citep[see e.g.][]{Li12}. There is a convergence of the cluster volume occupied by magnetic fields of the highest magnitudes ($|\vect{B}|>10^{-4}$\,G): $\sim1\%$ for '125:Base', $\sim5\%$ for '250:Base', and $\sim7\%$ for '500:Base' case (see Fig.~\ref{fig:filfac}). Hence, we could estimate that in higher resolution simulations (requiring > $10^7$ CPU-hours) this value is going to converge under $\sim10\%$.

\subsection{Effects of thermal conduction}
\label{sec:tc}

\begin{figure*}
	\includegraphics[scale=0.3]{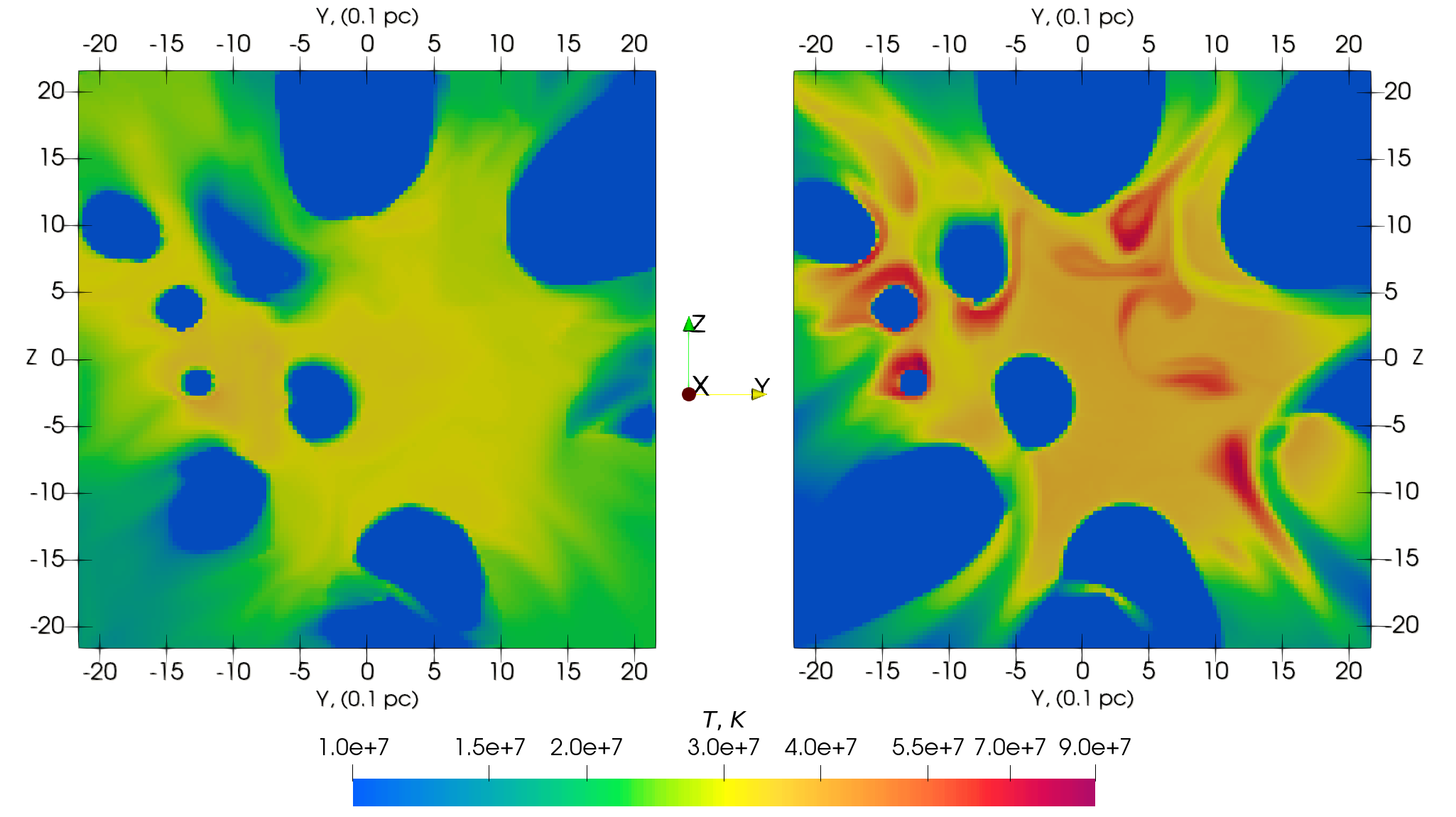}
	\vspace*{-1mm}
    \caption{The effect of plasma thermal conduction on the temperature maps of the inner part of the cluster. Left panel: the map constructed with account of the thermal conduction model described in $\S$ \ref{sec:tc}. Right panel: the map simulated within ideal MHD model without the thermal conduction.}
    \label{fig:tc}
\end{figure*}

\begin{figure}
	\includegraphics[width=8.5cm]{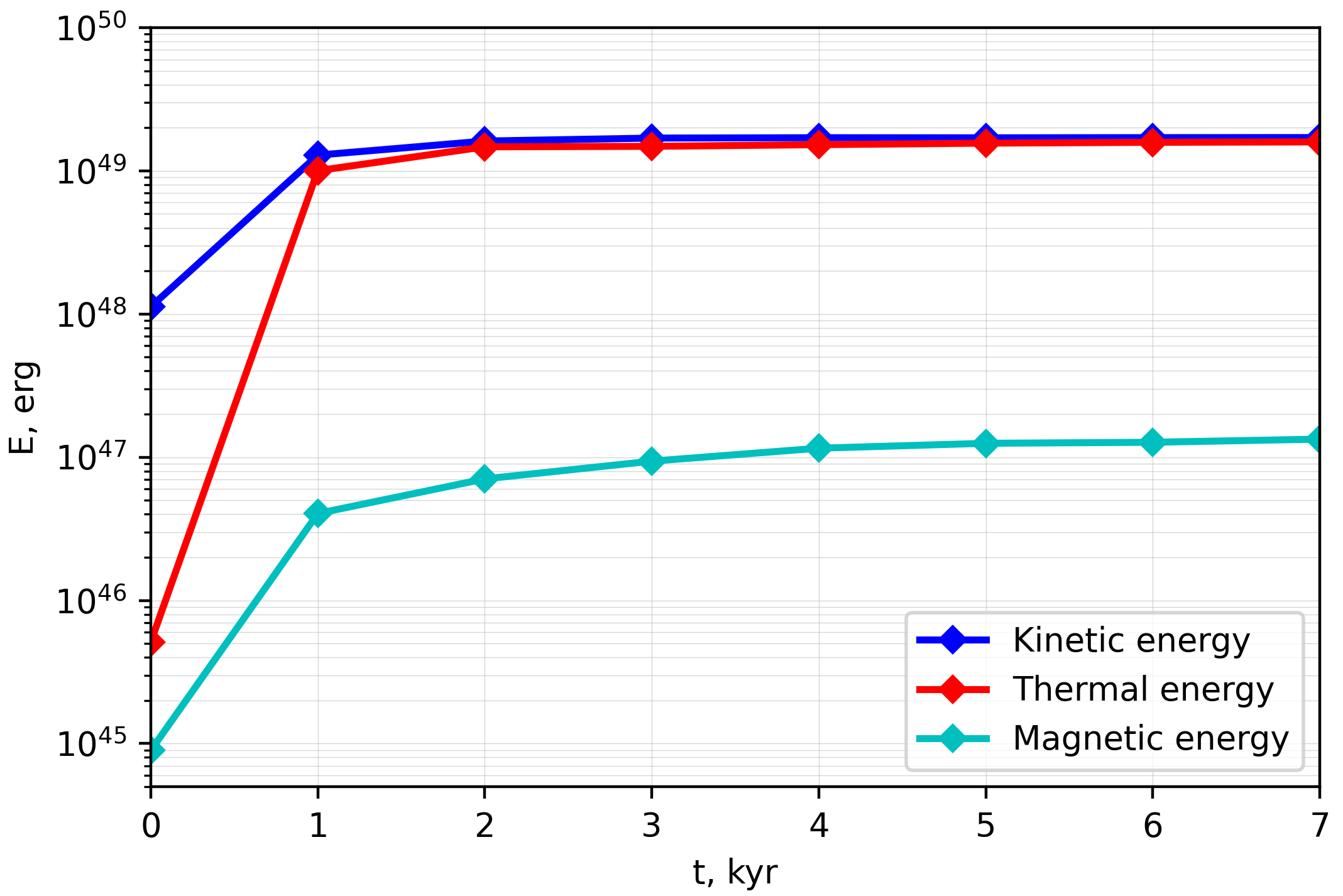}
	\vspace*{-1mm}
    \caption{Total kinetic, thermal, and magnetic energies deposited in the computational domain as functions of the integration time.}
    \label{fig:enpart}
\end{figure}

Here we present the results of the '125:Base' and '125:Base-TC' simulations. The first one assumes an inefficient electron heating, thus the heat conduction is negligible. The second one, in contrast, speculates on the electron-ion temperature equilibration, which results in an effective heat conduction regime. In Fig.~\ref{fig:tc} one could clearly see the difference between the temperature maps of the two mentioned simulations. Effective thermal conduction allows the heat to be evenly distributed through the cluster core volume. As a consequence, the overheated areas in the proximity of strongly suppressed O-winds effectively disappear, so the typical temperature in the cluster core falls down to $\sim3\times10^{7}$\,K with fairly weak and smooth temperature variation across the thermalised medium of the cluster. For some regions we also report a significant change in the geometry of the plasma flows when the thermal conduction is acting.

\section{Discussion}
\label{sec:dis}
The main new results of the present MHD simulations of the inner core of the YMSC at the stage with the active WR and O-star winds are related to the spatial structure and the magnitudes of the magnetic fields and the thermalisation of the winds kinetic power with an account of the thermal conduction.

\subsection{Magnetic field amplification in the cluster core}
\label{sec:MF}
Besides heating plasma in the inner part of the YMSC, the colliding fast supersonic winds of massive stars are carrying magnetic fields which are rooted at the surface of rotating  stars as it is the case of Parker's model of solar wind \citep{Par58}. The models of the heliospheric magnetic fields in the solar wind were widely discussed \citep[see e.g.][and the references therein]{Axf72,Zank22} and see \citet[][]{Herbst22} for a recent review of models of stellar astrospheres. In particular, it was shown by \citet[][]{Axf72} and \citet[][]{Cra74} that in nearly incompressible subsonic flow downstream of the spherical wind termination shock the azimuthal magnetic field $B_{\phi}$ would increase as $\propto{r}$ up to the magnitude which is dynamically significant to affect the plasma flow. In the solar wind the effect implies that a magnetic ridge is produced in the heliopause near the sub-stagnation point \citep[e.g.][]{Ner93}. They also pointed out the role of the magnetic field reconnection of alternating magnetic polarity strips in the heliopause.

Since the magnetic fields injected by the young massive stars into the intercluster plasma are anisotropic, we simulated in this paper the plasma flows and magnetic fields in clusters with both randomly oriented and aligned stellar spins. The rotation axis directions of the massive cluster members may not be necessarily distributed isotropically. The strong stellar spin alignment was reported in 48 stars by \citet[][]{spin17} from two old open clusters NGC 6791 and NGC 6819. The authors argued that the alignment likely holds from the cluster formation moment and is caused by the global angular momentum of the cluster-forming cloud. In young massive stellar cluster R136 in LMC \citet[][]{R136rot12} measured a rotation of the cluster as a whole which is supporting indirectly the idea of the possible alignment origin. This idea is currently under discussion as the projected inclinations of stars measured by \citet[][]{spin21} in the Pleiades and Praesepe clusters are consistent with both moderate alignment and isotropic spins, while the isotropic stellar spin orientations were suggested for NGC 2516 \citep{spin20}. The stellar spin orientation distribution could affect the magnetic field structure in YMSCs. However, in our magnetic field maps simulated for the randomly oriented as well as for the aligned stellar spins we found no significant qualitative differences between the two configurations neither in the magnetic field topology nor in the amplitude of the amplified magnetic fields. 

In Figs. \ref{fig:MF1}, \ref{fig:mfperp} and \ref{fig:3dmf} the projection and rendering maps of the magnetic fields in the inner region of a young stellar cluster simulated with 3D MHD model are shown.
The intermittent structure of magnetic fields with large-scale field filaments of magnitude well exceeding $100$\,$\mu$G is apparent. The origin of the strong magnetic field amplification can be associated with the local action of the mentioned above Axford-Cranfill effect downstream of the termination shocks of the powerful stellar winds of massive stars. The high thermal pressure in the central region of the cluster, produced by the colliding stellar winds, regulates the sizes of the wind bow shocks and helps to compress the magnetic fields in the filaments. Large magnitudes of magnetic fields are also apparent in the regions of the fast winds collisions as it is illustrated in Figs. \ref{fig:MF1} and \ref{fig:mfperp}. While the non-linear ideally conductive MHD simulations consider the possible dynamical effect of the fields on the flow, the magnetic reconnection effects cannot be accounted for in our model. Despite the relatively large magnitudes of the magnetic fields in filaments, the plasma parameter $\beta$ (see $\S$~\ref{sec:eleq}) is typically larger than 10 over the inner cluster domain. The presence of relatively high large-scale magnetic fields in the stellar clusters with multiple shocks provides favourable conditions for high-energy particle acceleration in these systems. 
Indeed, as one can see in Fig.~\ref{fig:mfperp} that magnetic fields are amplified to magnitudes $\sim0.3$\,mG by the colliding flows of powerful stellar winds in the cluster. The scale sizes of the high magnitude magnetic field filament shape structures are $l\sim 0.5$\,pc. A simple estimate of the maximal energy $\epsilon_m$ of particle accelerated in the system can be done using the Hillas approach as $\epsilon_m \sim u/clB$ \citep{Hillas84}, where the plasma bulk velocity field $u$ is shown in Fig.~\ref{fig:velo}. The estimation shows that the clusters can confine and accelerate protons to energies above $\gsim 100$\,TeV. The highly amplified magnetic fields are enveloping the regions with the strong termination shocks of the  fast stellar winds which can inject and accelerate non-thermal protons. To construct the energy spectra of accelerated particles kinetic simulations should be performed, which requires some model of magnetic turbulence in the dynamical range extending down to short scales, not resolved in the current 3D MHD simulations.

\subsection{Energy partitioning and thermalisation}

\begin{figure}
	\includegraphics[width=8.7cm]{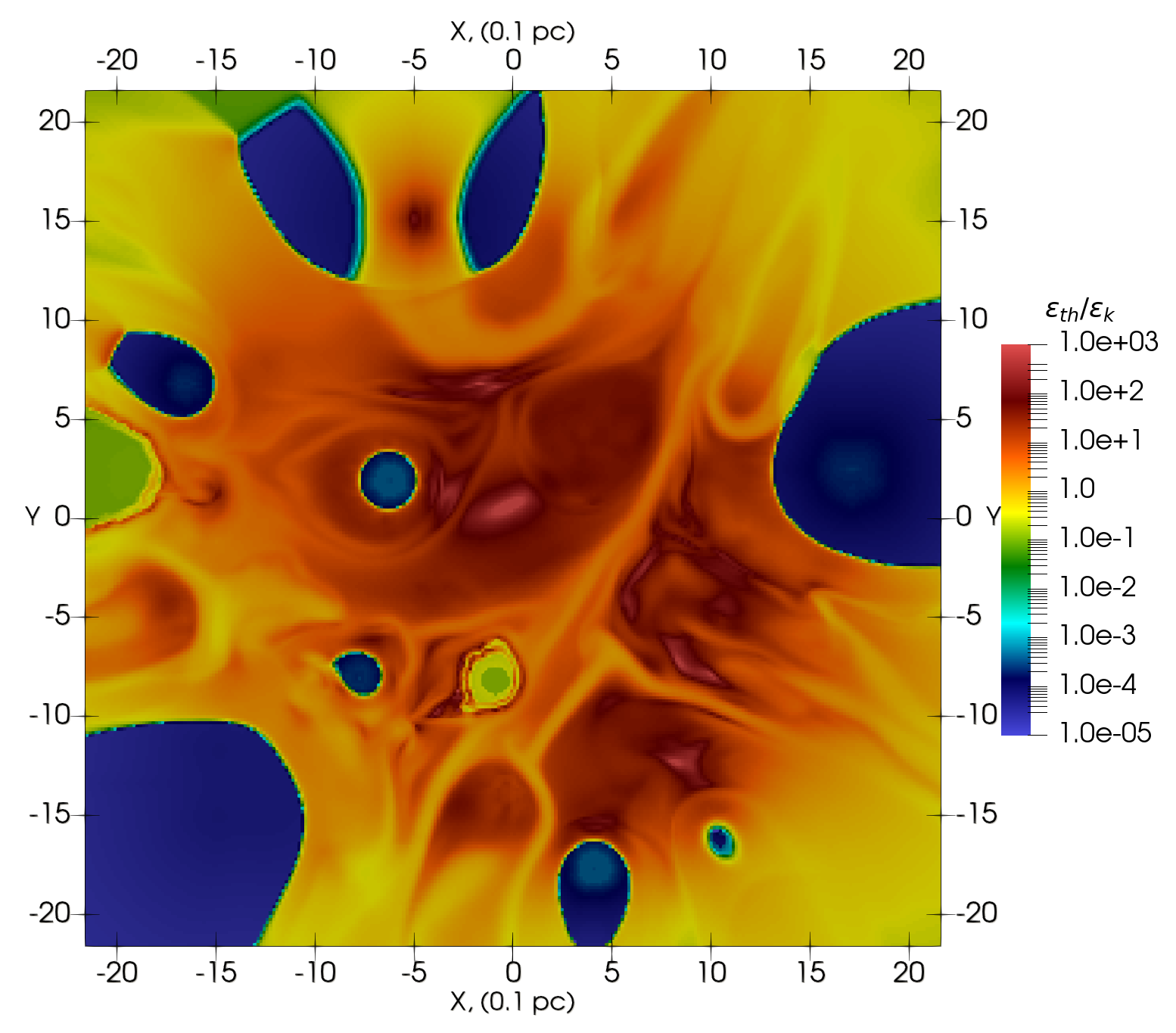}
	\vspace*{-5mm}
    \caption{Map of thermal to kinetic energy density ratio in $xy$-plane slice of $z=6$. Note the two CSG wind blown bubbles (light green areas), the first one near $(x,y)=(-1,-7)$ and the second one at the left edge, around $y=2.5$.}
    \label{fig:thermal}
\end{figure}

The colliding winds are effectively thermalised in the cluster core. In the 'Base' case the total mechanical luminosity supply from 60 objects is $\dot{E}_{\mathrm{kin}}\sim1.15\times10^{39}$\ergs. The quasi-stationary regime of the energy partition and flow geometry in the computational domain is reached after about 4 kyr of the integration time, see Fig.~\ref{fig:enpart}. After the establishing of quasi-stationary flow regime the total energy budget captured inside the domain is $E_{\mathrm{tot}}\sim3.3\times10^{49}$\,erg, including: $E_{\mathrm{kin}}\sim1.7\times10^{49}$\,erg, $E_{\mathrm{th}}\sim1.6\times10^{49}$\,erg, and $E_{\mathrm{mf}}\sim1.5\times10^{47}$\,erg. Thus $E_{\mathrm{th}}$/$E_{\mathrm{tot}}$ is $\sim48\%$, giving us the energy distribution inside the cluster close to the equipartition between thermal and kinetic energy. This ratio is preserved in the remaining models, including the one with the thermal conduction ('Base-TC'). In Fig.~\ref{fig:thermal} we present the map of the local kinetic energy density to thermal energy density ratio, which has the highest values $\sim10^{2}$-$10^{3}$ closer to the cluster centre. The energy flow across the domain boundaries is characterised by a slightly different distribution: $\dot{E}_{\mathrm{th(out)}}$/$\dot{E}_{\mathrm{tot(out)}}$ is $\sim36\%$ for 'Base' and other models, except for the 'Base-TC', where $\dot{E}_{\mathrm{th(out)}}$/$\dot{E}_{\mathrm{tot(out)}}$ is $\sim46\%$. 
Taking into account that the cluster has almost only mechanical energy supply in the form of fast stellar winds, these results imply an effective thermalisation inside the cluster. We also conclude that the thermalisation efficiency does not depend on the $\dot{E}_{\mathrm{kin}}$ and radius $r$ of the cluster (i.e. the compactness).

\subsection{X-ray diffuse emission of the hot plasma in the cluster core}
\label{sec:them}

\begin{figure}
\includegraphics[width=8.5cm]{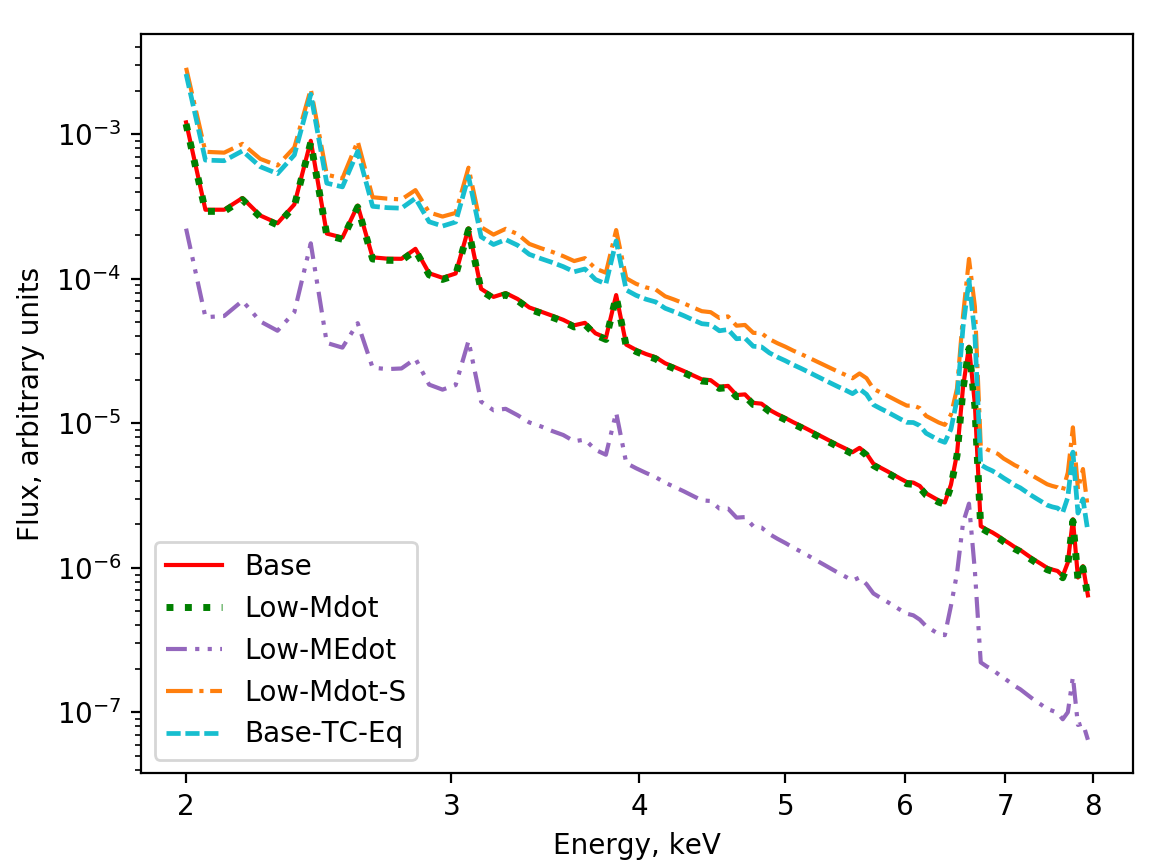}
\vspace*{-1mm}
\caption{X-ray spectra for the chosen set of models. Here, 'Base' is the main model (see the parameters in Table \ref{tab:tab1}).
'Low-Mdot' is the same as 'Base', but with reduced RSG/YSG masses, which gives the same X-ray spectrum, while 'Low-MEdot' has reduced O/WR masses (and therefore, reduced kinetic energy release). 
'Low-Mdot-S' is the same as 'Low-Mdot', but within the 1 pc domain. Finally, 'Base-TC-Eq' is calculated assuming the electron-ion temperature equilibration with  $T_{\mathrm{e}} = T_{\mathrm{eq}} = (T_{\mathrm{i}} + 0.1 T_{\mathrm{i}})/2$ and thermal conduction turned on.}
\label{fig:allspec}
\end{figure}

\begin{table}
\centering
\caption{Obtained effective temperatures and 2-8 keV luminosities for the chosen set of models.}
\label{tab:tab3}
    \begin{tabular}{lcr}
	    \hline
		 Model & $T_{\mathrm{eff}}$\,(keV) & $L$\,(\ergs) \\
        \hline
        	 Base & 1.25 & $4.28\times10^{33}$\\
		 Low-Mdot & 1.25 & $4.19\times10^{33}$ \\
		 Low-MEdot & 1.12 & $7.27\times10^{32}$ \\
         Low-Mdot-S & 1.40 & $1.14\times10^{34}$ \\
		 Base-TC-Eq & 1.35 & $9.77\times10^{33}$ \\ 
		 \hline
   	\end{tabular}\label{XR}
\end{table}
  
The nature and the mechanisms of the X-ray diffuse emission of compact clusters are often obscure, allowing a freedom of interpretation (see $\S$ \ref{sec:intro}). That is why it is a relevant task to calculate the thermal X-ray spectra for our models of a cluster and analyse the differences in effective temperatures (here and below -- the temperature which gives the same form of thermal spectrum as the summarized spectrum of a model cluster) and X-ray luminosities within the set of models. We took the energy range of 2-8 keV for building the illustrative spectra.

From the \textmc{pluto} simulations we obtained the values of ion temperatures and densities in every cell of the model cluster. To find the corresponding electron temperatures in each cell, we used the Spitzer equation solution \citep[][]{Spi62} with respect to the electron heating time, i.e. approximate time $t_{\mathrm{ex}}$ the emitting matter needs to leave the emission region. The initial electron temperature is taken as $\sim 0.1$ of the ion temperature as expected in the material, accelerated on the strong shock (see $\S$ \ref{sec:eleq}). To find $t_{\mathrm{ex}}$, in the \textmc{pluto} simulation we followed the transport of the labelled piece of matter from the centre of the cluster. It turned out that it takes $\sim 5,000$ yrs to leave the 2 pc cluster and $\sim 2,500$ yrs to leave the 1 pc cluster. This time-scale is reduced for the matter closer to the edge of cluster. Then we divided the cluster on several layers and for each layer used its own $t_{\mathrm{ex}}$ according to the distance to the centre – from 5,000 yrs to 1,000 yrs for 2 pc cluster and from 2,500 yrs to 500 yrs for 1 pc cluster. Solving the Spitzer equation for each cell, we found the time $\Delta t$, during which the electrons from that cell are heated enough to contribute to the 2-8 keV thermal emission (we put this cutoff on $T_0=0.8$ keV). The value $\Delta t/t_{\mathrm{ex}}$ represents the fraction of cluster volume containing heated electrons. Thus, we multiplied the emission power on the coefficient $\Delta t/t_{\mathrm{ex}}$ to consider the fact that electrons have a finite lifetime in the cluster and need time to heat. The final electron temperature for the given cell is taken as Spitzer equation solution for a moment $t$ between $t_0$, when $T_{\mathrm{e}}$ firstly becomes higher than $T_0$, and $t_{\mathrm{ex}}$, when the emitting matter should leave the domain: $t=(t_0+t_{\mathrm{ex}})/2$. With obtained this way distribution of $T_{\mathrm{e}}$ and known density distribution it is possible to calculate the thermal X-ray emission spectrum from the cluster core. To do so, we used the \textmc{apec} spectral model \citep[][]{SmithApec} from the \textmc{xspec} 12.12.0 package  \citep[][]{ArnaudXSPEC}. 

In addition to the approach described above, we examined the X-ray spectrum in the limiting case of the electron-ion temperature equilibration and the presence of thermal conduction, as discussed in $\S$ \ref{sec:thcon2}. In that case $T_{\mathrm{e}} = T_{\mathrm{eq}} = (T_{\mathrm{i}} + 0.1 T_{\mathrm{i}})/2$. Here we did not take into account the time of the electron heating as discussed before, considering that the equilibration is obtained in turbulence dissipation processes. Therefore, the 'Base-TC-Eq' model with equilibration is close to what we could get if the \textmc{pluto} simulations were two-fluid, including electrons and ions.
	
In Fig.~\ref{fig:allspec} spectra for models 'Base', 'Low-Mdot', 'Low-MEdot', 'Low-Mdot-S' and 'Base-TC-Eq' are shown. The first four of model spectra allow examining the impact of source compactness ('Low-Mdot-S' vs 'Low-Mdot'), mechanical energy input ('Base' vs 'Low-MEdot') and the presence of dense RSG/YSG envelopes ('Base' vs 'Low-Mdot') on thermal emission. The last one allows studying how the presence of thermal conduction and the assumption of electron-ion temperature equilibration ('Base' vs 'Base-TC-Eq') affects the thermal spectrum. The effective temperatures and X-ray luminosities for all five models are shown in Table \ref{XR}. One can see that the reduction of the cluster radius with the same $\dot{E}_{\mathrm{cl}}$ and $\dot{M}_{\mathrm{cl}}$ leads to a significant increase in X-ray luminosity. This is due to the density increase in the compact cluster and, therefore, more efficient heating of the electrons. The decrease of the cluster mechanical luminosity, as expected, inevitably leads to lower X-ray luminosity. Enlarging the mass of dense envelopes does not affect the emission, because their temperatures are much lower than needed for emission in 2-8 keV range. Finally, the electron-ion temperature equilibration combined with thermal conduction gives us higher effective emission temperature. Also a slightly higher level of X-ray luminosity is obtained.

3D MHD modeling of the interacting winds of massive stars among other things opens good possibilities for investigating the nature of X-ray diffuse emission of young compact clusters. X-ray luminosities in 2-8 keV energy range obtained for our models vary from $\sim7\times10^{32}$\ergs to $\sim1\times10^{34}$\ergs, which is in a good agreement with \citet[][]{RP14}, despite their modeling is quite different from the one presented in this paper. In the frame of 3D HD model they found that X-ray luminosity of a cluster at the WR winds stage is $\sim10^{34}$\ergs, then it drops to $\sim10^{32}$\ergs and remains so until the moment of supernova explosion in the cluster which greatly increase the X-ray luminosity.
What is important, the known young massive stellar clusters also have their observed diffuse X-ray emission luminosities \citep[see Table~3 in][]{RP14} in the range close to our predictions. As for the effective temperatures, it is hard to compare our results with the ones, obtained by fitting the observed YMSC spectra, as the latter most often have two or three thermal components, or additional non-thermal component.

Given the scale sizes of the compact clusters and the angular resolution of X-ray instruments it is possible to obtain the maps and spectra of the inner part of the cluster after the subtraction of the numerous point sources mostly associated with the hot atmospheres of massive stars. Our simulations predict some global morphology of the hot plasma in the cluster core which may be useful for analysis of the high resolution \textit{Chandra} and \textit{XMM-Newton} data. For example, we present in Fig.~\ref{fig:temp} a map of the plasma temperature simulated by the single fluid code. This map clearly shows some departures from the spherical symmetry in the outer regions of the simulated cluster which can be tested with a reasonably long \textit{Chandra} observation. For example, Westerlund 1 observations with \textit{Chandra} provided data for a few azimuthal sectors from 1’ to 5’ from the center of the cluster \citep[][]{Mun06}. The fluxes for the external boundary of the simulated cluster can be compared with the fluxes of a few different azimuthal segments within the annulus between 1’-2’ in Westerlund 1. The X-ray spectra simulations for the inner part of the cluster also allow a comparison with existing  observations of the cluster cores spectra. Given the estimated distance to Westerlund 1 of $\sim 4$ kpc, one arcminute radius corresponds to about 1 pc. Our simulated spectra can be compared with the Chandra data from the central part of Westerlund 1, and,  within the uncertainties of the model which we discussed in the text, they are reasonably consistent. The individual modeling of the emission of known massive clusters at the evolutionary stage with powerful stellar winds on the basis of MHD \textmc{pluto} simulations will be presented in the separate paper.

\section{Conclusion}
\label{sec:con}
We have investigated the structure of plasma flows, magnetic fields and thermal X-ray emission from the inner core of a young massive star cluster at the evolutionary stage dominated by OB- and WR-star winds. There are a number of young stellar clusters at this evolutionary stage in the Galaxy -- Arches, Quintuplet, Westerlund 1 and 2 and the others \citep[][]{PZ10} which allow detailed multi-wavelength studies. YMSCs are considered to play an important role in the starburst galaxies \citep{Kru19}. 

With 3D single-fluid MHD modeling we have found that magnetic fields in the cluster core have filamentary structure. The filaments with typical scale sizes of $\sim$ 0.5 pc and field magnitudes reaching $\sim$ 300 $\mu$G are enveloping the most powerful stellar winds of WR and, especially, O stars. The magnetic structures are produced by the Axford-Cranfill type amplification mechanism in the subsonic outflows behind the stellar wind termination shocks. The amplified fields are further compressed by the high pressure hot gas produced in the central regions of the core by the colliding stellar winds. That is why the highly amplified fields are clearly seen in the regions between the colliding winds. 

The presence of the high magnetic field filaments in the parsec scale cluster core makes YMSCs favorable places of high-energy cosmic ray acceleration. This may help to understand the origin of multi-TeV gamma-ray emission from the vicinity of the compact stellar clusters Westerlund 1 and 2 recently reported in  \citep[][]{Aha19,Wd1_HESS22,MestreWd2} in the context of cosmic ray acceleration models \citep[see for a review][]{Byk20}.         

The thermal conduction and kinetic energy thermalisation effects are essential for modeling of plasma flows dynamics and radiation of YMSCs. While the single-fluid 3D MHD simulations allow only some parametric studies of the effects caused by electrons, we modeled the limiting cases of the strong and, vice versa, absent thermal conduction. The results show that the presence of thermal conduction affects significantly the temperature distribution inside the cluster core and, therefore, the thermal energy flux across the borders of the domain. The ratio of the thermal to kinetic energy both inside the domain and flowing outside shows an effective thermalisation of the stellar winds mechanical energy in the cluster. This is important \citep[][]{Chv85,SH03} for detailed modeling of the cluster wind formation outside the inner cluster core.  

The X-ray diffuse emission of the cluster cores can be resolved with current generation of the X-ray  observatories. The X-ray spectra of hot plasma depend on the mechanical energy supply by young massive stars, mass loading and the effects of thermal conduction. While the form of the spectrum remains quite similar for all the models studied here with only marginal differences, the X-ray luminosities derived for models with different energy input, cluster core size and thermal conduction may vary noticeably. They are generally in a good correspondence with the observations of galactic YMSCs. Our modeling of the magnetic field structure provides a base for upcoming studies of the diffuse non-thermal emission from the inner cores of young massive star clusters.

\section*{Acknowledgements}
We wish to thank the referee for a careful reading of our paper and constructive comments. This research made use of \textmc{pluto} public MHD code developed by A. Mignone and the \textmc{pluto} team. We acknowledge the use of of data provided by NASA ADS system and SIMBAD database, operated at CDS, Strasbourg, France.  Some of the modeling was performed at the JSCC RAS and the 'Tornado' subsystem of the St.~Petersburg Polytechnic University super-computing centers. D.V.B. acknowledge a support from RSF 21-72-20020 grant.

\section*{Data Availability}

The output data may be provided upon a reasonable request. 


\bibliographystyle{mnras}
\bibliography{mnras_template} 


\bsp	
\label{lastpage}
\end{document}